\documentclass[letterpaper, 10 pt, conference]{ieeeconf}  % Comment this line out
\IEEEoverridecommandlockouts                              % This command is only
\usepackage{graphicx}
\usepackage{float}
\usepackage{booktabs}
\usepackage{array}
\usepackage{amsmath} % assumes amsmath package installed
\usepackage{url}
\usepackage{amsmath}
\makeatletter
\def\maketag@@@#1{\hbox{\m@th\normalfont\normalsize#1}}
\makeatother
\usepackage{algorithm}
\usepackage[noend]{algpseudocode}
\usepackage{setspace}
\usepackage[keeplastbox]{flushend}
\usepackage{tikz}
\usepackage{lipsum}
\usepackage{comment}
\usepackage{amsfonts}
\usepackage{amssymb}
\usepackage[normalem]{ulem}
\usepackage{multirow}
\usepackage{subfigure}
\usepackage{cite}
\usepackage{fancyhdr}

\fancypagestyle{firstpage}{%
  \lhead{\vspace{-0.75 cm} \small \textbf{{\fontfamily{cmss}\selectfont
2019 Annual American Control Conference (ACC)\\July 10–-12, 2019, Philadelphia, PA, USA}}}
}

\title{\LARGE \bf
Energy Management for Autonomous Underwater Vehicles using Economic Model Predictive Control
}

\author{Niankai Yang, Dongsik Chang, Mohammad Reza Amini, Matthew Johnson-Roberson, and Jing Sun  
\thanks{Niankai Yang, Dongsik Chang, Mohammad Reza Amini, Matthew Johnson-Roberson, and Jing Sun are with Department of Naval Architecture and Marine Engineering, University of Michigan, Ann Arbor, MI 48109, USA. (e-mail: \{ynk,\;dongsik,\;mamini,\;mattjr,\;jingsun\}@umich.edu)}% ,   
}
\begin{document}

\maketitle
\thispagestyle{firstpage}

\makeatletter
\AfterEndEnvironment{algorithm}{\let\@algcomment\relax}
\AtEndEnvironment{algorithm}{\kern2pt\hrule\relax\vskip3pt\@algcomment}
\let\@algcomment\relax
\newcommand\algcomment[1]{\def\@algcomment{\footnotesize#1}}
\renewcommand\fs@ruled{\def\@fs@cfont{\bfseries}\let\@fs@capt\floatc@ruled
  \def\@fs@pre{\hrule height.8pt depth0pt \kern2pt}%
  \def\@fs@post{}%
  \def\@fs@mid{\kern2pt\hrule\kern2pt}%
  \let\@fs@iftopcapt\iftrue}
\makeatother

%%%%%%%%%%%%%%%%%%%%%%%%%%%%%%%%%%%%%%%%%%%%%%%%%%%%%%%%%%%%%%%%%%%%%%%%%%%%%%%%%%%%
\begin{abstract}
This paper investigates the problem of energy-optimal control for autonomous underwater vehicles (AUVs). To improve the endurance of AUVs, we propose a novel energy-optimal control scheme based on the economic model predictive control (MPC) framework. We first formulate a cost function that computes the energy spent for vehicle operation over a finite-time prediction horizon. Then, to account for the energy consumption beyond the prediction horizon, a terminal cost that approximates the energy to reach the goal (energy-to-go) is incorporated into the MPC cost function. %
%%%%%%%%%%%%%%%%%%%%%%%%%%%%%%%
To characterize the energy-to-go, a thorough analysis has been conducted on the globally optimized vehicle trajectory computed using the direct collocation (DC) method for our test-bed AUV, \textit{DROP-Sphere}. %
%%%%%%%%%%%%%%%%%%%%%%%%%%%%%%%
Based on the two operation modes observed from our analysis, the energy-to-go is decomposed into two components: (i) \textit{dynamic} and (ii) \textit{static} costs. This breakdown facilitates the estimation of the energy-to-go, improving the AUV energy efficiency. Simulation is conducted using a six-degrees-of-freedom dynamic model identified from \textit{DROP-Sphere}. The proposed method for AUV control results in a near-optimal energy consumption with considerably less computation time compared to the DC method and substantial energy saving compared to a line-of-sight based MPC method. 
\end{abstract}
% developed by the DROP Lab at the University of Michigan
% To characterize the energy-to-go, a thorough analysis has been conducted on the solution obtained from a global optimization method for our test-bed AUV, \textit{DROP-Sphere}.
%%%%%%%%%%%%%%%%%%%%%%%%%%%%%%%%%%%%%%%%%%%%%%%%%%%%%%%%%%%%%%%%%%%%%%%%%%%%%%%%%%%%
%%%%%%%%%%%%%%%%%%%%%%%%%%%%%%%%%%%%%%%%%%%%%%%%%%%%%%%%%%%%%%%%%%%%%%%%%%%%%%%%%%%%
\section{INTRODUCTION}

The autonomous underwater vehicle (AUV) has become an essential tool for long-range and deepwater underwater missions. The scope of the missions is often restricted by the endurance of AUVs. To improve AUV endurance, several approaches through design modification have been proposed (e.g.,~\cite{webb2001slocum,santhakumar2013power}). However, considering the limited internal space and desired performance requirements of the vehicle, the improvement in AUV endurance achieved through design modification is limited, thereby motivating the development of energy-optimal control methods.

The existing literature on the energy-optimal control of AUVs focuses primarily on utilizing the ocean currents in the planning stage. 
In \cite{garau2005path}, the $A^*$ algorithm was used to compute a horizontal energy-minimum path in a 2D ocean environment with eddies by assuming that the energy cost is equivalent to the traveling time. The genetic algorithm was employed in~\cite{alvarez2004evolutionary} to calculate the optimal path in a spatio-temporally varying current environment using a kinematic model for predicting the energy consumption. However, due to the expensive computations involved in these planning algorithms \cite{luo2013improved}, optimal paths have to be computed offline. Execution-level controllers are then utilized to handle the internal and external uncertainties during real-time execution of the planned optimal paths.

The energy-optimal control in the execution stage has also been studied in \cite{sarkar2016modelling,geranmehr2015nonlinear}. Given the planned references computed in the planning stage, these energy-optimal execution-level controllers achieve energy reduction by minimizing a weighted sum of control efforts and path tracking error. However, the energy saving potential of these control strategies decreases with the increase in the path tracking error. Our previous work \cite{yang2018real} studied the energy-optimal control problem for the horizontal straight-line motion of AUVs, where we proposed an execution-level controller based on model predictive control (MPC). The proposed controller addresses the real-time trajectory optimization and energy management of AUVs without considering steering and diving maneuvers.

In this paper, we generalize the previous approach in \cite{yang2018real} to the horizontal motion of an AUV that involves steering by developing a controller based on economic model predictive control (EMPC) \cite{ellis2014tutorial}. The proposed EMPC formulation consists of energy-dependent stage and terminal costs that take into account the energy spent over the prediction horizon and that beyond the prediction horizon (energy-to-go). To capture the energy-to-go associated with the optimal maneuver, we analyze the velocity and thrust profiles %
%%%%%%%%%%%%%%%%%%%%%%%%%%%%%%%
obtained by optimizing the global vehicle trajectory using the direct collocation (DC) method. %
%%%%%%%%%%%%%%%%%%%%%%%%%%%%%%%
Based on the analysis, we decompose the vehicle trajectory beyond the prediction horizon into the \textit{dynamic} and \textit{static} stages to facilitate the estimation of the energy-to-go. The effectiveness of our control design is demonstrated through simulations.

% To capture the energy-to-go associated with the optimal maneuver, we analyze the velocity and thrust profiles of the vehicle operation achieved with a global optimization method (direct collocation) using a model derived for our test-bed AUV.

%The remainder of this paper is organized as follows: in Section \ref{section.2}, the specifications and model of our test-bed AUV is introduced. The AUV energy management problem is described in Section \ref{section.3}. Section \ref{section.4} proposes the overall EMPC framework. Simulation results and analysis are presented in Section \ref{section.5}. Finally, conclusions and future work are discussed in Section \ref{section.6}.

%%%%%%%%%%%%%%%%%%%%%%%%%%%%%%%%%%%%%%%%%%%%%%%%%%%%%%%%%%%%%%%%%%%%%%%%%%%%%%%%%%%% 
%%%%%%%%%%%%%%%%%%%%%%%%%%%%%%%%%%%%%%%%%%%%%%%%%%%%%%%%%%%%%%%%%%%%%%%%%%%%%%%%%%%%
\section{AUV Model} \label{section.2}
\subsection{DROP-Sphere Configuration}
\begin{figure}[!h]
\centering
\includegraphics[width=2.7in]{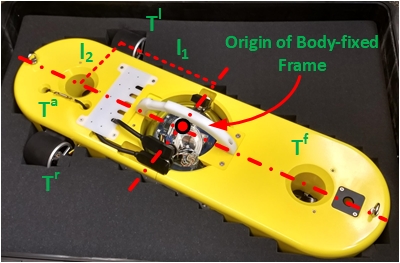} \vspace{-0.2cm}
\caption{Schematic of the DROP-Sphere}  
\label{fig:Sphere} \vspace{-0.6cm}
\end{figure}

In this study, the \textit{DROP-Sphere} platform, an open-source, low-cost, $6000\ m$ rated AUV developed by the DROP Lab at the University of Michigan \cite{iscar2018towards}, is used for controller development and demonstration. \textit{DROP-Sphere} has an elliptical body of $1\ m$ length and $0.5\ m$ width (see Fig.~\ref{fig:Sphere}). The weight and displacement of the vehicle are $20.42\ kg$ and $20.57\ kg$, respectively. The vehicle is equipped with four hub-less bi-directional thrusters for surge, heave, pitch, and yaw controls. A transparent sphere with other mechanical and electrical devices is placed at the center of the vehicle.  

\subsection{Mathematical Modeling}
\begin{figure}[!h]
\centering
\includegraphics[width=2.7in]{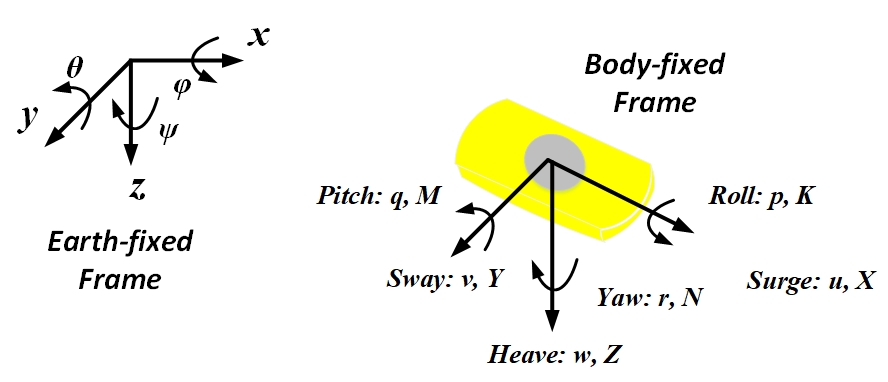} \vspace{-0.3cm}
\caption{Reference frames and notations} 
\label{fig:reference frames} \vspace{-0.4cm}
\end{figure}

To model the motion of \textit{DROP-Sphere}, we use a general motion model for AUVs developed in \cite{prestero2001verification}. The six-degrees-of-freedom (DOF) motion is described in a body-fixed coordinate frame and an earth-fixed coordinate frame (see Fig.~\ref{fig:reference frames}). The vector of velocities denoted by $\nu$, the vector of positions and orientations denoted by $\eta$, and the vector of external forces and moments denoted by $\tau$ are defined as \vspace{-0.1cm}
\begin{equation} \vspace{-0.1cm}
\nu = \left[                 
  \begin{array}{cccccc}   
    u & v & w & p & q & r\\  
  \end{array}
\right] ^T,
\end{equation}
\begin{equation} \vspace{-0.1cm}
\eta = \left[                 
  \begin{array}{cccccc}   
    x & y & z & \phi & \theta & \psi \\  
  \end{array}
\right] ^T,
\end{equation}
\begin{equation} \vspace{-0.1cm}
\tau = \left[                 
  \begin{array}{cccccc}   
    X & Y & Z & K & M & N \\  
  \end{array}
\right] ^T,
\end{equation}
where the components of $\nu$, $\eta$, and $\tau$ are illustrated in Fig.~\ref{fig:reference frames}. \vspace{-0.8cm}

With the defined notations, the kinematic model of an AUV, which relates the linear and angular velocities with the positions and orientations, is described as \vspace{-0.1cm}
\begin{equation}  \label{eq:Vehicle Kinematics}  \vspace{-0.1cm}
\dot{\eta} = J(\eta)\nu,
\end{equation}
where $J(\eta)$ is the coordinate transformation matrix. 

The dynamic model of an AUV establishes the relationship between external forces and vehicle states. The system dynamics is expressed as \vspace{-0.1cm}
\begin{equation}  \label{eq:Vehicle Dynamics}  \vspace{-0.1cm}
M_t\dot{\nu}+F_c(\nu)+F_h(\nu)\nu+F_g(\eta) = \tau_c,\\ 
\end{equation}  
where $M_t, F_c(\nu), F_h(\nu), F_g(\eta)$ and $\tau_c$ are the vehicle total mass, the Coriolis and centripetal forces, the hydrodynamic damping coefficients, the hydrostatic forces, and the control inputs. In this study, only the diagonal terms in $M_t$ and the diagonal and quadratic terms in $F_h(\nu)$  are considered.

Based on the thruster allocation of \textit{DROP-Sphere}, the control input vector $\tau_c$ is defined as \vspace{-0.3cm}

\begin{scriptsize} 
\begin{equation} \label{Eq:control_input_transformation} \vspace{-0.1cm}
\setlength{\arraycolsep}{2pt}
\tau_c =\left[                 
  \begin{array}{cccccc}   
    T^h & 0 & T^a+T^f & 0 &  T^hl_3 + l_1(T^a-T^f) & l_2(T^r-T^l)\\  
  \end{array}
\right] ^T,
\end{equation}
\end{scriptsize}%
where, as shown in Fig.~\ref{fig:Sphere}, $T^a, T^f, T^r, T^l$ are the aft, fore, right, and left thrusts, $l_1$ is the distance between vertical thrusters and the midship, and $l_2$ is the distance between horizontal thrusters and the center line. $T^h=T^l+T^r$ is the total horizontal thrust. $l_3$ is the vertical distance between horizontal thrusters and the center of gravity. To facilitate the energy consumption analysis, the power conversion relationship $P(T^i)$ that converts the force from $i^{th}$ ($i=\{a,f,r,l\}$) thruster to the power consumption is adopted from \cite{spangelo1994trajectory}. %

%%%%%%%%%%%%%%%%%%%%%%%%%%%%%%%%%%%%%%%%%%%%%%%%%%%%%%%%%%%%%%%%%%%%%%%%%%%%%%%%%%%%
%%%%%%%%%%%%%%%%%%%%%%%%%%%%%%%%%%%%%%%%%%%%%%%%%%%%%%%%%%%%%%%%%%%%%%%%%%%%%%%%%%%%
\section{Energy Management for AUVs} \label{section.3}
\subsection{Problem Formulation}
%The essential task of the AUV energy management at the execution stage is to regulate the individual thrusters to minimize the overall vehicle energy consumption subject to both local constraints (e.g., the vehicle dynamics and thruster limitation) and global constraints (e.g., the references generated during the planning stage)~\cite{musardo2005ecms}. 
Let us assume that the vehicle operates in an obstacle-free underwater environment without ocean currents. Define $\ell(T^l_k,T^r_k,T^f_k,T^a_k) = P(T^l_k)+P(T^r_k)+P(T^f_k)+P(T^a_k)$ as the power consumption of all four thrusters. Then, the energy management (EM) problem of an AUV is formulated as the following optimal control problem:\vspace{-0.4cm}

\begin{small}
\begin{equation}  \label{eq:EM-Cost-Function}  \vspace{-0.2cm}
\begin{split}
\mathop{\min}_{\{ T^l_{k} \},\{ T^r_{k} \},\{ T^f_{k} \},\{ T^a_{k} \}} &J(\chi_{0},\chi_f,\{ T^l_{k} \},\{ T^r_{k} \},\{ T^f_{k} \},\{ T^a_{k} \}) \\
&= \sum_{k=0}^{N_f-1}\ell(T^l_k,T^r_k,T^f_k,T^a_k)\Delta t , 
\end{split}
\end{equation}
\end{small}
subject to \vspace{-0.2cm}
\begin{equation} \label{eq:EM-Constraints} \vspace{-0.1cm}
\begin{split}
\chi_{k+1}=f&(\chi_{k},T^l_{k},T^r_{k},T^f_{k},T^a_{k}), \; \chi_{k} \in \mathbb{X}, \; \chi_{N_f} \in \mathbb{X}_f,\\
&T^l_{k} \in \mathbb{U}, \; T^r_{k} \in \mathbb{U},\;  T^f_{k} \in \mathbb{U},\; T^a_{k} \in \mathbb{U}, \\
\end{split}
\end{equation}
where $\chi = [\nu, \eta]$ is the states of the vehicle, $\chi_0$ and $\chi_f$ are the initial and the desired final states, $\{T^i_k\}$ is the input sequence from the $i$ thruster, $N_f$ is the total number of time steps, $\Delta t$ is the time step, and $f(\cdot)$ is the 6 DOF system dynamics discretized from (\ref{eq:Vehicle Kinematics}) and (\ref{eq:Vehicle Dynamics}). $\mathbb{X}$, $\mathbb{U}$ and $\mathbb{X}_f$ are the constraints for the states, inputs and final state.

In this study, we consider the vehicle operation between two horizontal waypoints, ($x_0$,$y_0$) and ($x_f$,$y_f$). The two waypoints are selected such that the initial vehicle heading ($\psi_0 = 0$) is not pointing towards the final position. Thus, steering maneuver is required to drive the vehicle to the destination. The initial surge velocity $u_0$ is set as the surge velocity that optimizes the total energy consumption when the yaw and pitch control energies are negligible \cite{yang2018real}. This selection of $u_0$ represents an optimal straight-line cruising condition of a vehicle before turning. The other velocities and orientations for $\chi_0$ are set to zero, and the velocities and orientations for $\chi_f$ are assumed to be unconstrained.

\subsection{Horizontal Motion Optimization} \label{Section.3B}

\begin{figure}[!h]
\centering
\includegraphics[width=3.0in]{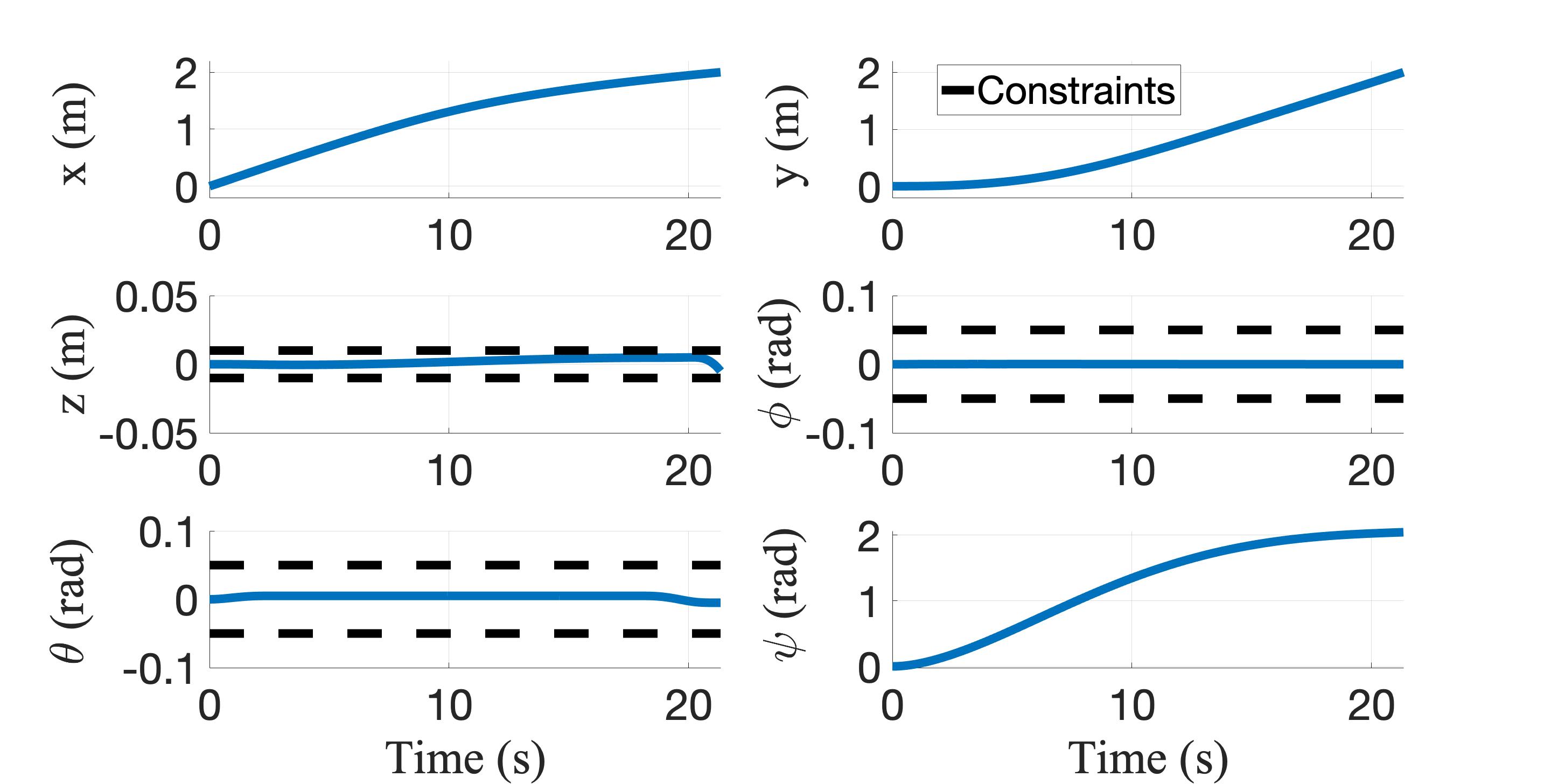} \vspace{-0.3cm}
\caption{Position and orientation traces from direct collocation} 
\label{fig:DC_trajectory} \vspace{-0.6cm}
\end{figure}

To solve the EM problem in (\ref{eq:EM-Cost-Function}) and (\ref{eq:EM-Constraints}), DC, a numerical method for solving the optimal control problems \cite{von1992direct}, can be employed 
to optimize the vehicle trajectory globally. %
For a specific example with ($x_0=0$, $y_0=0$) and ($x_f=2$, $y_f=2$), a discretization with 100 intervals is used when approximating the state and input trajectories with the trapezoid rule. The constraints include: $\mathbb{X}=\{|z| \le 0.01, |\phi| \le 0.05, |\theta| \le 0.05\}$, $\mathbb{X}_f=\{ (x-x_f)^2+(y-y_f)^2 \le 0.05^2 \}$ and $\mathbb{U}=\{|T| \le 7.86\}$. The units for the variables are $m$ for positions, $rad$ for orientations, and $N$ for the thrusts. The optimal trajectories resulted from DC are illustrated in Fig.~\ref{fig:DC_trajectory}. It can be seen from Fig.~\ref{fig:DC_trajectory} that the vehicle satisfies all the constraints and reaches $\mathbb{X}_f$ in $21.33\ s$. However, the computing time for DC to obtain this solution is $1181.04\ s$ on a 2.9 GHz Intel Core i5 processor with 8GB RAM. Thus, during the execution stage, the DC method can only be used as an open-loop control law, which makes it suffer from the robustness issue against uncertainties. \vspace{-0.3cm}

\begin{figure}[!h] 
\centering
\includegraphics[width=2.7in]{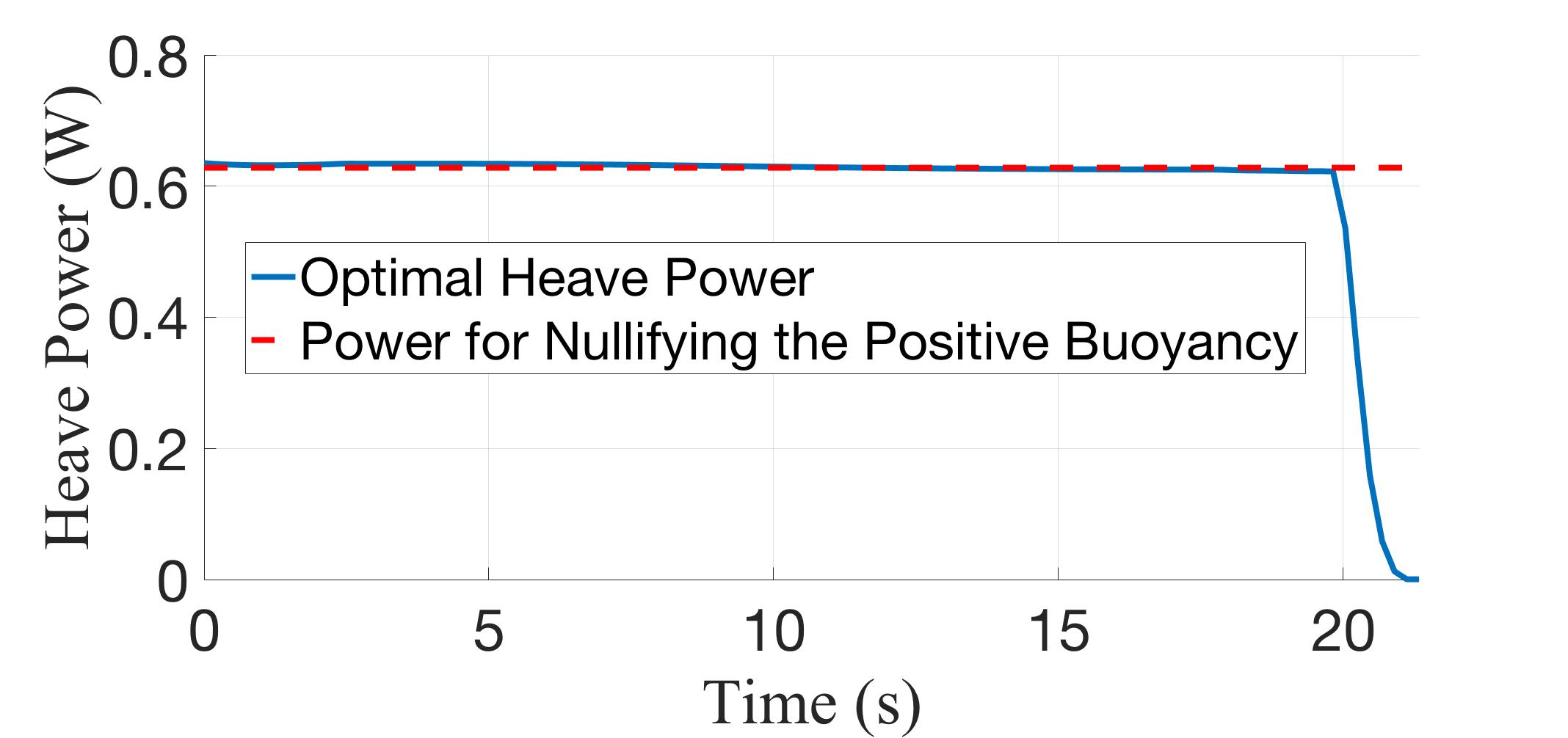} \vspace{-0.3cm}
\caption{Heave power history from direct collocation} 
\label{fig:Heave_power_history} \vspace{-0.3cm}
\end{figure}

As illustrated with the numerical example, intensive computation of the DC method prohibits it from real-time implementation, especially for resource-limited AUV platforms such as \textit{DROP-Sphere}. To facilitate the development of an execution-level controller, the optimal solution obtained from the DC method is analyzed to simplify the EM problem in (\ref{eq:EM-Cost-Function}) and (\ref{eq:EM-Constraints}).  The total energy consumption resulted from the DC method is distributed for surge ($33.52\%$), heave ($60.84\%$),  yaw  ($5.54\%$)  and  pitch  ($0.11\%$)  controls, which suggests that the pitch control energy is negligible. In addition, considering that the heave power is constant during most of the trajectory (see Fig.~\ref{fig:Heave_power_history}), the heave energy will be a function of the vehicle travel time in the horizontal plane. Then, pitch and heave controls can be decoupled from (\ref{eq:EM-Cost-Function}) and (\ref{eq:EM-Constraints}), and performed by PID controllers. Therefore, we can focus on the optimization problem for the horizontal motion given as 
\begin{equation}  \label{eq:Horizontal-optimization-Cost-Function} \vspace{-0.2cm}
\begin{split}
\mathop{\min}_{\{ T^l_{k} \},\{ T^r_{k} \}} &J_h(\zeta_{0},\zeta_f,\{ T^l_{k} \},\{ T^r_{k} \}\})  \\
&= \sum_{k=0}^{N_f-1}(\ell_h(T^l_k,T^r_k)+P^{PB})\Delta t, 
\end{split}
\end{equation}
subject to \vspace{-0.2cm}
\begin{equation} \label{eq:Horizontal-optimization-Constraints}  \vspace{-0.1cm}
\begin{split}
\zeta_{k+1}=f_h&(\zeta_{k},T^l_{k},T^r_{k}), \; \zeta_{N_f} \in \mathbb{X}_f,T^l_{k} \in \mathbb{U}, \; T^r_{k} \in \mathbb{U},\\
\end{split}
\end{equation}
where $\zeta$ = [$u,v,r,x,y,\psi$] and $\zeta_f$ are the state vector and the desired final state in the horizontal plane, $\ell_h(T^l_k,T^r_k) = P(T^l_k)+P(T^r_k)$ is the power consumption of two horizontal thrusters, $P^{PB}=2P(\frac{B-W}{2})$ is the power consumed by two vertical thrusters for nullifying the positive buoyancy \cite{yang2018real}, and $B$ and $W$ are the vehicle buoyancy and weight. The decoupled vehicle dynamics $f_h(\cdot)$ is obtained by assuming $z,\phi,\theta$ to be constant within unit time step. \vspace{-0.4cm}

%, and $\mathbb{Z}_f=\{ (x_k-x_f)^2+(y_k-y_f)^2 \le 0.05^2 \}$ is the final state constraint
%%%%%%%%%%%%%%%%%%%%%%%%%%%%%%%%%%%%%%%%%%%%%%%%%%%%%%%%%%%%%%%%%%%%%%%%%%%%%%%%%%%%
%%%%%%%%%%%%%%%%%%%%%%%%%%%%%%%%%%%%%%%%%%%%%%%%%%%%%%%%%%%%%%%%%%%%%%%%%%%%%%%%%%%%

\section{EMPC for AUV Energy Management} \label{section.4}
\subsection{EMPC Formulation}

To solve the optimization problem of horizontal motion in \eqref{eq:Horizontal-optimization-Cost-Function} and \eqref{eq:Horizontal-optimization-Constraints}, we can design a controller based on the energy consumption of the thrusters as the stage cost under the EMPC framework. However, this EMPC formulation will not drive the vehicle to the destination unless the remaining energy to reach the destination (energy-to-go) is included into the optimization. Thus, a terminal cost reflecting the energy-to-go is proposed, which leads to the following EMPC formulation
\begin{equation} \label{eq:EMPC-Objective-function} 
\begin{split}
&\mathop{\min}_{\{ T^l_{k|t} \},\{ T^r_{k|t} \}} J_{EMPC}(\zeta_{0},\zeta_f,\{ T^l_{k|t} \},\{ T^r_{k|t} \}) \\
&= \sum_{k=0}^{N-1}(\ell_h(T^l_{k|t}, T^r_{k|t})+P^{PB})\Delta t+K_h(\zeta_{N|t},\zeta_f), 
\end{split}
\end{equation}
subject to \vspace{-0.2cm}
\begin{equation} \label{eq:EMPC-Constraints} \vspace{-0.1cm}
\begin{split}
\zeta_{k+1|t}=f_h&(\zeta_{k|t},T^l_{k|t},T^r_{k|t}), \; T^l_{k|t} \in \mathbb{U}, \; T^r_{k|t} \in \mathbb{U},\\
\end{split}
\end{equation}
where $N$ is the prediction horizon, $(\cdot)_{k|t}$ is the $k$-step ahead prediction made at time instant $t$. According to the Bellman's Equation, %~\cite{bellman2013dynamic}, 
the optimal solution from solving (\ref{eq:EMPC-Objective-function}) and (\ref{eq:EMPC-Constraints}) will be equivalent to that of (\ref{eq:Horizontal-optimization-Cost-Function}) and (\ref{eq:Horizontal-optimization-Constraints}) if and only if the following equality holds \vspace{-0.1cm}
\begin{equation} \label{eq:c2g} \vspace{-0.1cm}
K_h(\zeta_{N|t},\zeta_f) = J_h^*(\zeta_{N|t},\zeta_f,\{ T^l\}^* ,\{ T^r \}^* ),
\end{equation}
where $J^*_h$ denotes the minimal energy consumption to drive the vehicle from $\zeta_{N|t}$ to $\zeta_f$ subject to the thruster input constraints. $\{ T^i\}^*$ ($i=\{r,l\}$) is the corresponding optimal thrust sequence. Therefore, in order to obtain a near-optimal energy consumption from the EMPC, a terminal cost, which approximates the optimal energy-to-go ($J^*_h$), is required. 

\subsection{Two-stage Energy-to-go Approximation}

\begin{figure}[!h]
\centering
\includegraphics[width=2.7in]{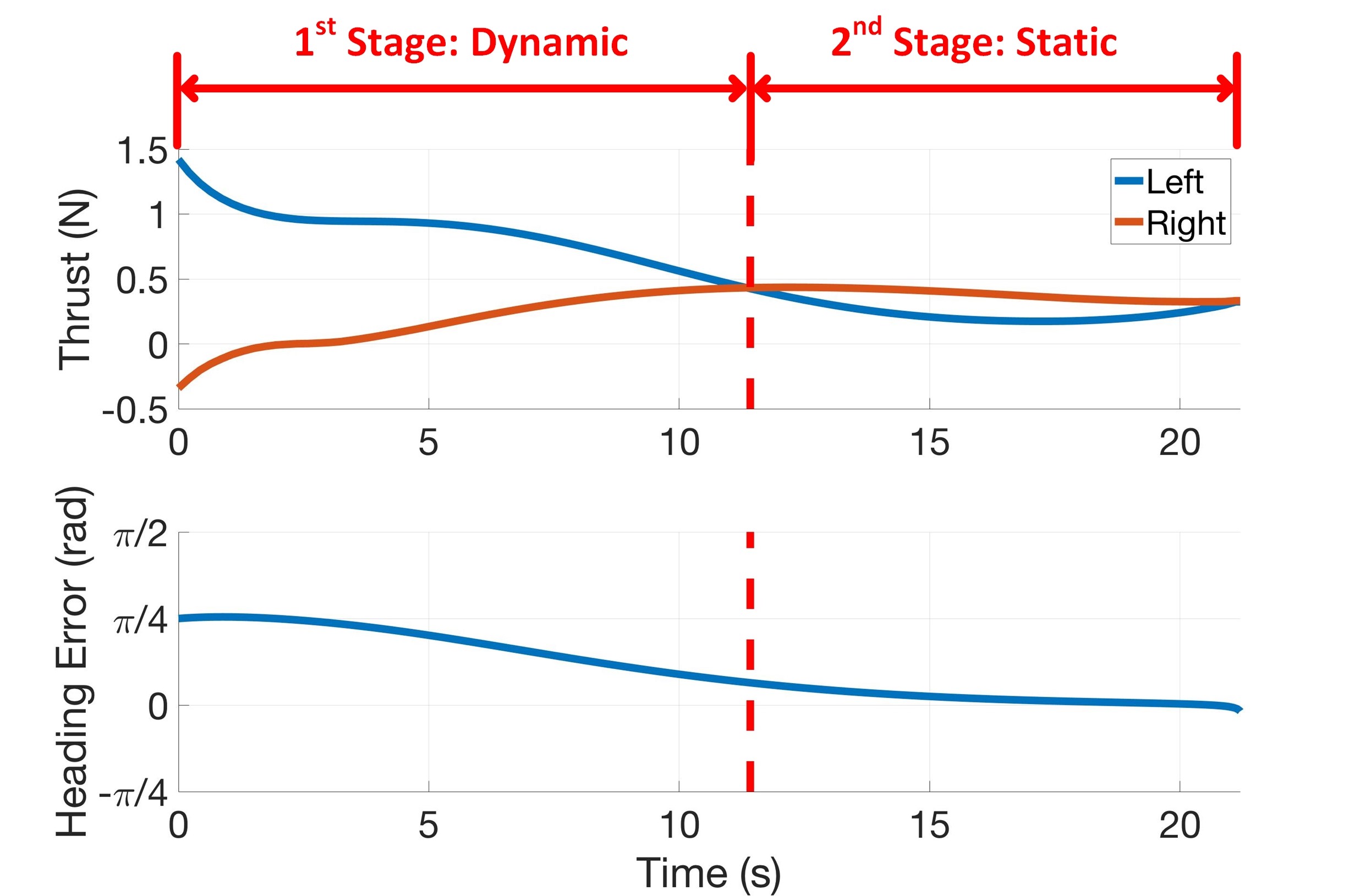}  \vspace{-0.3cm}
\caption{Thrusts and heading error from direct collocation}   \vspace{-0.7cm}
\label{fig:Horizontal Thruster Inputs from DC} 
\end{figure}

To approximate the optimal energy-to-go, an extensive analysis is conducted on the DC solution for understanding the characteristics of the optimal maneuver of \textit{DROP-Sphere}. The optimal input sequences from two horizontal thrusters and the heading error are shown in Fig.~\ref{fig:Horizontal Thruster Inputs from DC} for the case study presented in Section.~\ref{Section.3B}. The heading error, which represents the difference between current vehicle heading and the desired vehicle heading, is defined as \vspace{-0.1cm}
\begin{equation} \vspace{-0.1cm}
\Delta \psi_k = \tan^{-1}\frac{y_f-y_k}{x_f-x_k}-\tan^{-1}\frac{v_k}{u_k}-\psi_k.
\end{equation}
where $\tan^{-1}\frac{y_f-y_k}{x_f-x_k}$ is the path-tangential angle between vehicle and the final position, and $\tan^{-1}\frac{v_k}{u_k}$ is the drift angle. As illustrated in Fig.~\ref{fig:Horizontal Thruster Inputs from DC}, the operation of the vehicle can be divided into two stages, during which the difference between the two thrusters in the first stage is larger than that in the second stage. Considering the transformation between the thrusts and the resulted forces and moments in \eqref{Eq:control_input_transformation}, a larger difference in horizontal thrusts leads to a larger yaw moment for steering the vehicle, which is demonstrated by a larger decrease in heading error during the first stage in Fig.~\ref{fig:Horizontal Thruster Inputs from DC}.

Based on the characteristics of the DC solution, we separate the vehicle trajectory beyond the prediction horizon into dynamic and static stages by introducing an intermediate state. Then, a two-stage energy-to-go approximation problem can be formulated as\vspace{-0.3cm}

\begin{small}
\begin{equation} \label{eq:Two-stage_c2g_Cost} \vspace{-0.1cm}
K_h(\zeta_{N|t},\zeta_f) = \mathop{\min}_{u_s,v_s,r_s,x_s,y_s} (J_d(\zeta_{N|t},\zeta_s) + J_s(\zeta_s,\zeta_f)),
\end{equation}
\end{small}
subject to \vspace{-0.1cm}
\begin{equation} \label{eq:Two-stage_c2g_Constraints} \vspace{-0.1cm}
\begin{split}
&\zeta_{k+1|t}=f_h(\zeta_{k|t},T^l_{k|t},T^r_{k|t}), \; T^l_{k|t} \in \mathbb{U}, \; T^r_{k|t} \in \mathbb{U},\\
&\Delta \psi_s = \tan^{-1}\frac{y_f-y_s}{x_f-x_s}-\tan^{-1}\frac{v_s}{u_s}-\psi_s \approx 0.
\end{split}
\end{equation}
where $\zeta_s$ = [$u_s,v_s,r_s,x_s,y_s,\psi_s$] is the intermediate state with $J_d$ and $J_s$ being the dynamic and the static cost, respectively. The optimization of the intermediate state facilitates the optimal energy-to-go approximation by adjusting the durations of the dynamic and static stages.
%Note that, $J_d(\zeta_{N|t},\zeta_s)\approx J_h^*(\zeta_{N|t},\zeta_s,\{ T^l\}^* ,\{ T^r \}^* )$, and $J_s(\zeta_s,\zeta_f)\approx J_h^*(\zeta_s,\zeta_f,\{ T^l\}^* ,\{ T^r \}^* )$.

\subsubsection{Static Cost}

Our previous analysis in \cite{yang2018real} suggests that the total energy consumption mainly consists of heave and surge control energies when the yaw moment is negligible. Based on this analysis, we propose the following model to capture the energy use in the static stage \vspace{-0.1cm}
\begin{equation} \label{eq:static_cost} \vspace{-0.1cm}
J_s(u_s,v_s,x_s,y_s,\zeta_f) = \frac{d}{\sqrt{u_s^2+v_s^2}} (P^S + P^{PB}),
\end{equation}
where $d=\sqrt{(x_f-x_s)^2+(y_f-y_s)^2}$ is the horizontal distance between $\zeta_s$ and $\zeta_f$, and $P^S = 2P(\frac{X_{|u|u}|u_s|u_s}{2})$ is the power consumption from two horizontal thrusters. $X_{|u|u}$ is the damping coefficient for the surge motion.

\subsubsection{Dynamic Cost}

For the dynamic cost, in addition to the surge and heave control energies, the energy spent for yaw control should be considered. Since the horizontal thrusters are used for both surge and yaw controls, the surge and yaw control energies can be obtained by summing up the energy induced from the horizontal thrusters. To estimate the energy from the two horizontal thrusters, we consider the following surge and yaw equations of motion to parameterize the horizontal thrust sequences \vspace{-0.1cm}
\begin{equation} \label{eq:surge_EOM} \vspace{-0.1cm}
\begin{split}
(m-X_{\dot{u}})\dot{u} =& -(-vr+wq+Z_gpr)m-X_{|u|u}|u|u \\
&-(W-B)\sin\theta+T^l+T^r,
\end{split}
\end{equation}
\begin{equation} \label{eq:yaw_EOM} \vspace{-0.1cm}
\begin{small}
(I_{z}-N_{\dot{r}})\dot{r} = -(I_{y}-I_{x})pq-N_{|r|r}|r|r+l_2(T^r-T^l),
\end{small}
\end{equation}
where $m$ is the vehicle mass, $X_{\dot{u}}$ and $N_{\dot{r}}$ are the added mass, $N_{|r|r}$ is the hydrodynamic damping for yaw, $z_g$ is the center of gravity in the $z$ direction, and $I_{x}$, $I_{y}$ and $I_{z}$ are the moment of inertia in $x$, $y$ and $z$ directions. Then, by omitting the terms in (\ref{eq:surge_EOM}) and (\ref{eq:yaw_EOM}) that have negligible effect on surge and yaw dynamics during turning, following expressions are derived to approximate the optimal horizontal thrusts \vspace{-0.1cm}
\begin{equation} \label{eq:approximated_input_sequence} \vspace{-0.1cm}
\begin{cases}
    \tilde{T}^l_{k_1} =\frac{1}{2} (X_{|u|u}u^2_{k_1}+\frac{(I_{z}-N_{\dot{r}})\dot{r}_{k_1}+N_{|r|r}|r_{k_1}|r_{k_1}}{l_{2}}) ,\\
    \\
    \tilde{T}^r_{k_1} = \frac{1}{2} (X_{|u|u}u^2_{k_1}-\frac{(I_{z}-N_{\dot{r}})\dot{r}_{k_1}+N_{|r|r}|r_{k_1}|r_{k_1}}{l_{2}}),
  \end{cases}
\end{equation}
where $\tilde{T}^l_{k_1}$ and $\tilde{T}^r_{k_1}$ are the approximated optimal inputs for the left and right thrusters at time instant $k_1$, and $t_{k_1} = k_1 \Delta t$ represents the time within the dynamic stage.

\begin{figure}[!h]
\centering
\includegraphics[width=3.0in]{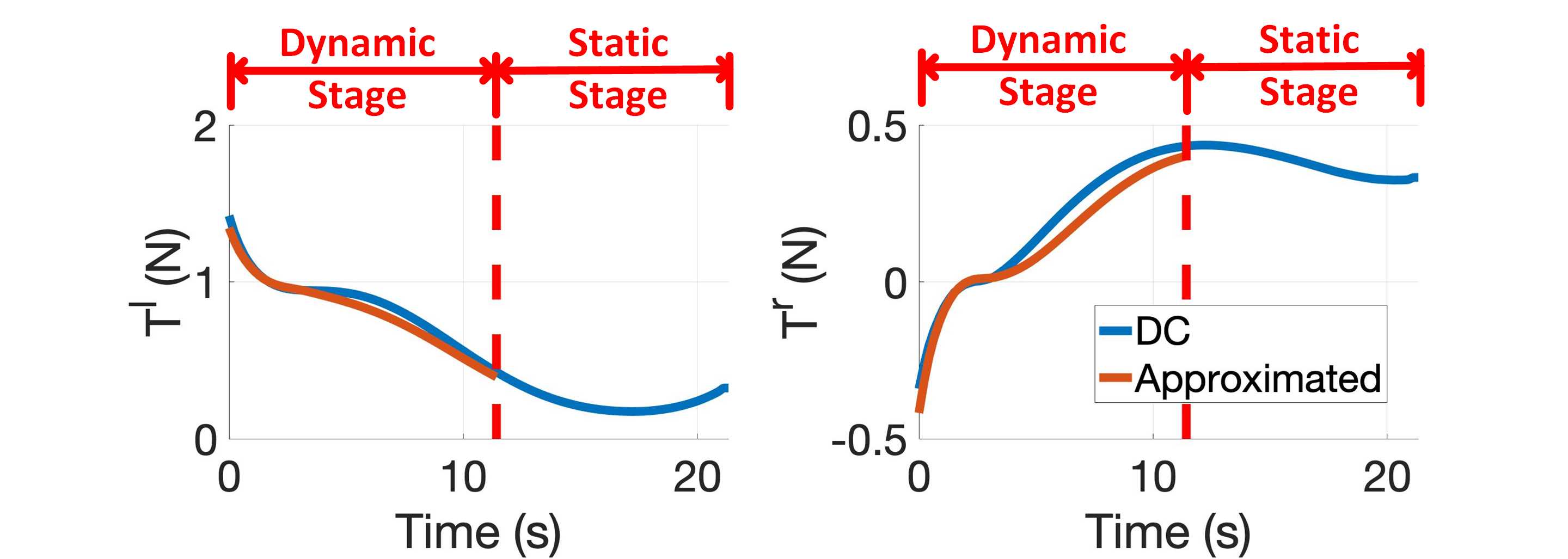} \vspace{-0.3cm}
\caption{Comparison of the approximated and optimal horizontal thrusts}  
\label{fig:Horizontal_Input_Comp} \vspace{-0.3cm}
\end{figure}

To validate the approximation in (\ref{eq:approximated_input_sequence}), $\{\tilde{T}^l_{k_1}\}$ and $\{\tilde{T}^r_{k_1}\}$ evaluated with the $u_{k_1}$, $r_{k_1}$, and $\dot{r}_{k_1}$ from the DC method are compared to the optimal thrust sequences obtained with the DC method (Fig.~\ref{fig:Horizontal_Input_Comp}). It can be seen from Fig.~\ref{fig:Horizontal_Input_Comp} that $\{\tilde{T}^l_{k_1}\}$ and $\{\tilde{T}^r_{k_1}\}$ properly capture the optimal patterns of the horizontal thrusters. \vspace{-0.2cm}

%To facilitate the parameterization of the horizontal thrusts, .

\begin{figure}[!h]
\centering
\includegraphics[width=3.0in]{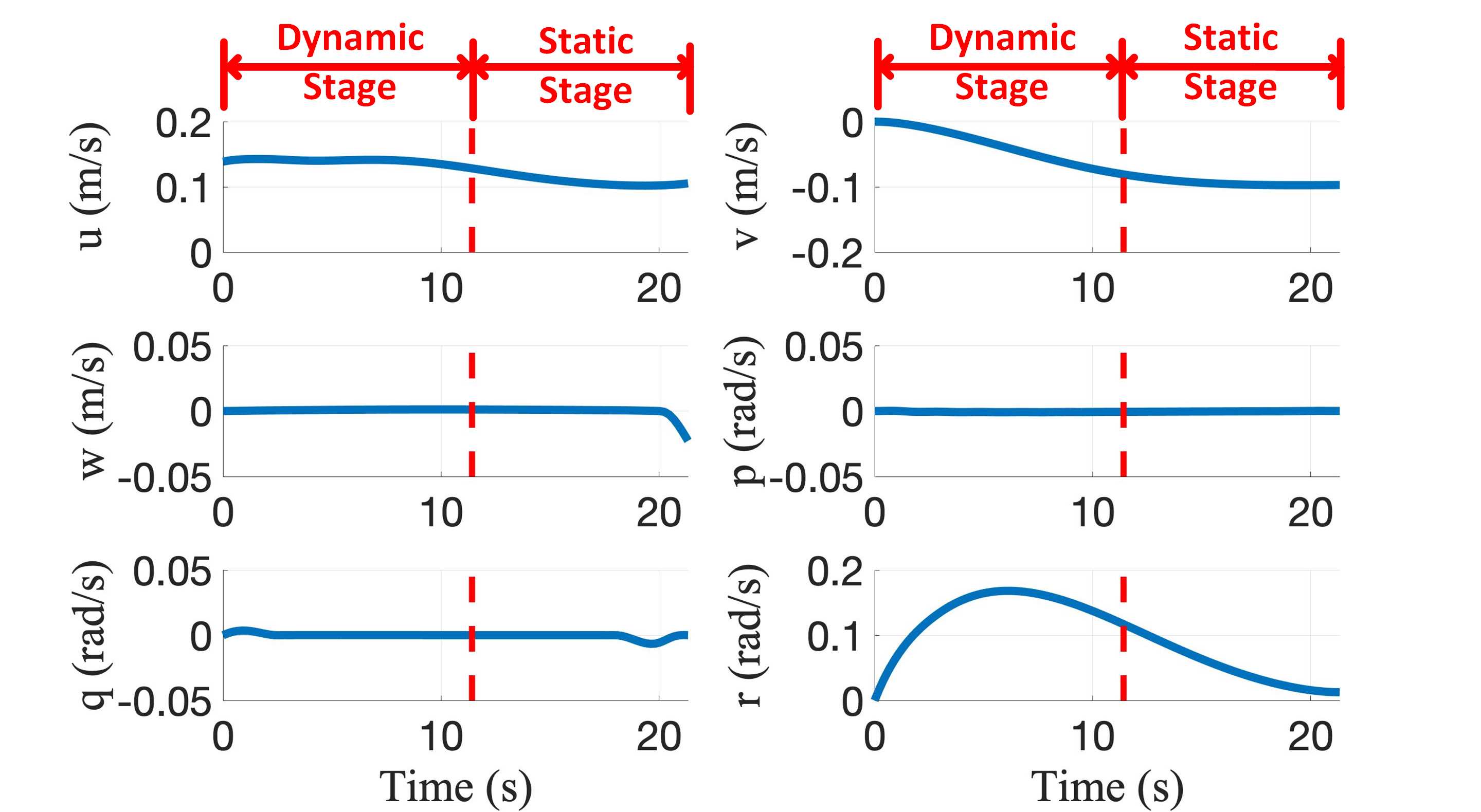} \vspace{-0.3cm}
\caption{Velocity traces from direct collocation} 
\label{fig:DC_velocities} \vspace{-0.3cm}
\end{figure}

To further simplify the dynamic cost, additional assumptions are made based on the velocity traces from the DC solution given in Fig.~\ref{fig:DC_velocities}: the surge velocity is assumed to be constant, and the yaw velocity is modeled as  \vspace{-0.1cm}
\begin{equation} \label{eq:approximated_yaw_velocity}
\tilde{r}_{k_1} = \begin{cases}
        r_{N|t}+\frac{4(\frac{\psi_d}{t_d}-r_{N|t})}{t_d}t_{k_1}, & \text{for } 0 \le t_{k_1} \le \frac{t_d}{2}\\
        \\
        \frac{\psi_d}{t_d}, & \text{for } t_{k_1} > \frac{t_d}{2}
        \end{cases}
\end{equation}
where $\tilde{r}_{k_1}$ is the approximated yaw velocity, $t_d$ is the duration of the dynamic stage, and $\psi_d = \psi_s - \psi_{N|t}$ is the heading variation within the dynamic stage. \vspace{-0.3cm}

With these assumptions, the dynamic cost can be described as %\vspace{-0.1cm}
% %
% \begin{equation} 
% J_d(\zeta_{N|t},\psi_d, t_d) = \sum_{k_1=N}^{N_s-1}(L_h(\tilde{T}^l_{k_1}, \tilde{T}^r_{k_1}) + P^{PB})\Delta t,
% \end{equation}
% %
%
\begin{equation} \label{eq:dynamic_cost}
J_d(\zeta_{N|t},\psi_d, t_d) = (P^h + P^{PB}) \cdot t_d,
\end{equation}
where $P^h = \frac{1}{4}(P(\tilde{T}^l_{t_0})+P(\tilde{T}^r_{t_0})+P(\tilde{T}^l_{t_d/2})+P(\tilde{T}^r_{t_d/2}))+\frac{1}{2}(P(\tilde{T}^l_{t_d})+P(\tilde{T}^r_{t_d}))$ is the estimated power consumption of the horizontal thrusters based on (\ref{eq:approximated_input_sequence}) and (\ref{eq:approximated_yaw_velocity}). 

\subsection{Energy-optimal EMPC}
With the estimated static and dynamic costs, the energy-to-go approximation problem in (\ref{eq:Two-stage_c2g_Cost}) and (\ref{eq:Two-stage_c2g_Constraints}) becomes
\begin{equation} \label{eq:Re-Two-stage_c2g_Cost}
\begin{split}
K_h(\zeta_{N|t},\zeta_f,\psi_d, t_d) = &\mathop{\min}_{t_d,\psi_d} (J_d(\zeta_{N|t},\psi_d, t_d) \\
& + J_s(u_s,v_s,x_s,y_s,\zeta_f)),
\end{split}
\end{equation}
subject to \vspace{-0.2cm}
\begin{equation} \label{eq:Re-Two-stage_c2g_Constraints}  \vspace{-0.1cm}
\begin{split}
\Delta \psi_s = \tan^{-1}\frac{y_f-y_s}{x_f-x_s}-\tan^{-1}\frac{v_s}{u_s}-\psi_s \approx 0.
\end{split}
\end{equation}
Note that the approximated optimal thrust sequences in the dynamic stage are functions of $t_d$ and $\psi_d$. Then, by integrating the system dynamics with the approximated optimal thrust sequences, the intermediate state will also depend on $t_d$ and $\psi_d$. However, obtaining an analytical expression by integrating the dynamics is nontrivial. Therefore, we assume $\psi_d = 2\Delta \psi_{N|t}$. Then, if we replace $\psi_s$ in \eqref{eq:Re-Two-stage_c2g_Constraints} with $\psi_d+\psi_{N|t}$ according to the definition of $\psi_d$, the $\Delta \psi_s$ will converge to zero as $\Delta \psi_{N|t}$ approaches zero. In addition, by assuming $u_{k_1}=u_{N|t}$, $v_{k_1}=v_{0|t}$ and $r_{k_1} = \frac{2\Delta \psi_{N|t}}{t_d}$, the horizontal position of the intermediate state can be derived based on the vehicle kinematic relationship in \eqref{eq:Two-stage_c2g_Constraints}
\begin{equation} \label{eq:horizontal_pos_intermediate_state}
\begin{cases}
    x_s = x_{N|t}+\frac{u_tt_d}{\Delta \psi_{N|t}}\sin(\Delta \psi_{N|t}) \cos(\psi_{nf}),\\
    \\
    y_s = y_{N|t}+\frac{u_tt_d}{\Delta \psi_{N|t}}\sin(\Delta \psi_{N|t}) \sin(\psi_{nf}),
  \end{cases}
\end{equation}
where $u_t = \sqrt{u^2_{N|t}+v^2_{0|t}}$ is the forwarding velocity, and $\psi_{nf}$ is the path-tangential angle between $\zeta_{N|t}$ and $\zeta_{f}$.

\begin{figure}[!h]
\centering
\includegraphics[width=2.7in]{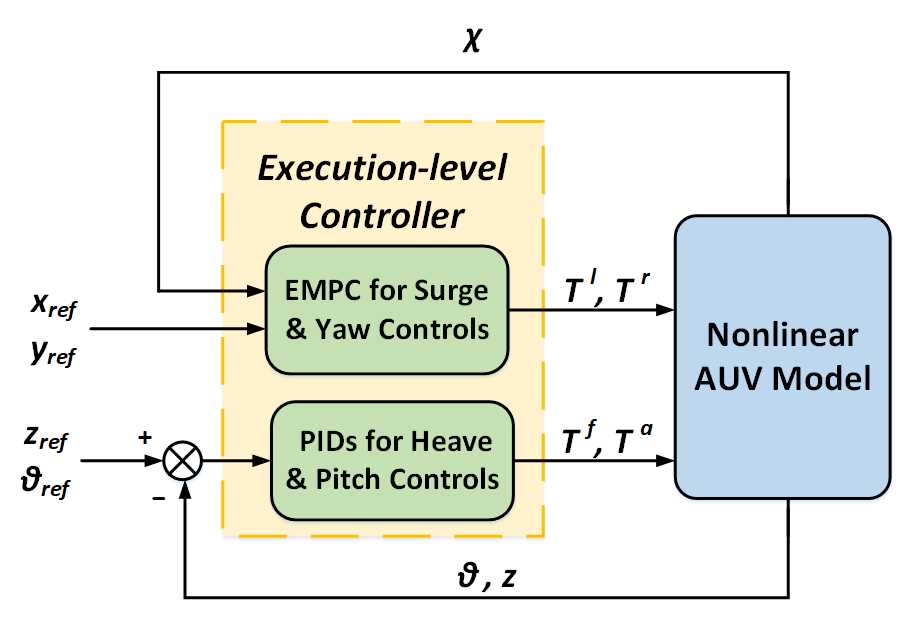} \vspace{-0.3cm}
\caption{Control architecture for EO-EMPC}  \vspace{-0.2cm}
\label{fig:EO_MPC_controller_architecture}
\end{figure}

Substituting the approximated intermediate state in \eqref{eq:horizontal_pos_intermediate_state} into \eqref{eq:Re-Two-stage_c2g_Cost}, the terminal cost $K_h$ is simplified as a function of the state at the end of the prediction horizon ($\zeta_{N|t}$), the duration of the dynamic stage ($t_d$) and the desired final state ($\zeta_f$). Then, based on the EMPC formulation in \eqref{eq:EMPC-Objective-function} and \eqref{eq:EMPC-Constraints}, the energy-optimal EMPC (EO-EMPC) is formulated as 
\begin{equation} \label{eq:EMPC-Objective-function_final} 
\begin{split}
\mathop{\min}_{\{ T^l_{k|t} \},\{ T^r_{k|t} \},t_d} J_{EO}(&\zeta_{0},\zeta_f,\{ T^l_{k|t} \},\{ T^r_{k|t} \})  \\
= \sum_{k=0}^{N-1}&(\ell_h(T^l_{k|t},T^r_{k|t})+P^{PB})\Delta t+ \\
&J_d(\zeta_{N|t},t_d)+J_s(\zeta_{N|t},\zeta_f,t_d), 
\end{split}
\end{equation}
subject to \vspace{-0.2cm}
\begin{equation} \label{eq:EMPC-Constraints_final}  \vspace{-0.1cm}
\begin{split}
\zeta_{k+1|t}=f_h&(\zeta_{k|t},T^l_{k|t},T^r_{k|t}), \; T^l_{k|t} \in \mathbb{U}, \; T^r_{k|t} \in \mathbb{U}.\\
\end{split}
\end{equation}
The overall schematic of EO-EMPC along with PID controllers for heave and pitch controls is given in Fig.~\ref{fig:EO_MPC_controller_architecture}. \vspace{-0.4cm}

%%%%%%%%%%%%%%%%%%%%%%%%%%%%%%%%%%%%%%%%%%%%%%%%%%%%%%%%%%%%%%%%%%%%%%%%%%%%%%%%%%%%
%%%%%%%%%%%%%%%%%%%%%%%%%%%%%%%%%%%%%%%%%%%%%%%%%%%%%%%%%%%%%%%%%%%%%%%%%%%%%%%%%%%%
\section{Simulation Results and Analysis} \label{section.5}
To verify the effectiveness of the proposed predictive controller, EO-EMPC is demonstrated through simulations using the six DOF model of \textit{DROP-Sphere} in MATLAB/Simulink. The sampling time of EO-EMPC ($\Delta t$) is $0.1\ s$, and the prediction horizon is $0.5\ s$. The performance of EO-EMPC is compared with those obtained from (i) direct collocation (DC) and (ii) line-of-sight-based MPC (LOS-MPC) for three case scenarios. LOS-MPC tracks the reference surge velocity and yaw angle under the standard MPC formulation (i.e., minimize the tracking error). The reference yaw angle is computed using the line-of-sight guidance law in \cite{lekkas2013line} with a lookahead distance of $0.5\ m$. The reference surge velocity is set as the initial surge velocity of the vehicle.

\begin{figure}[!h]
\centering
\includegraphics[width=2.5in]{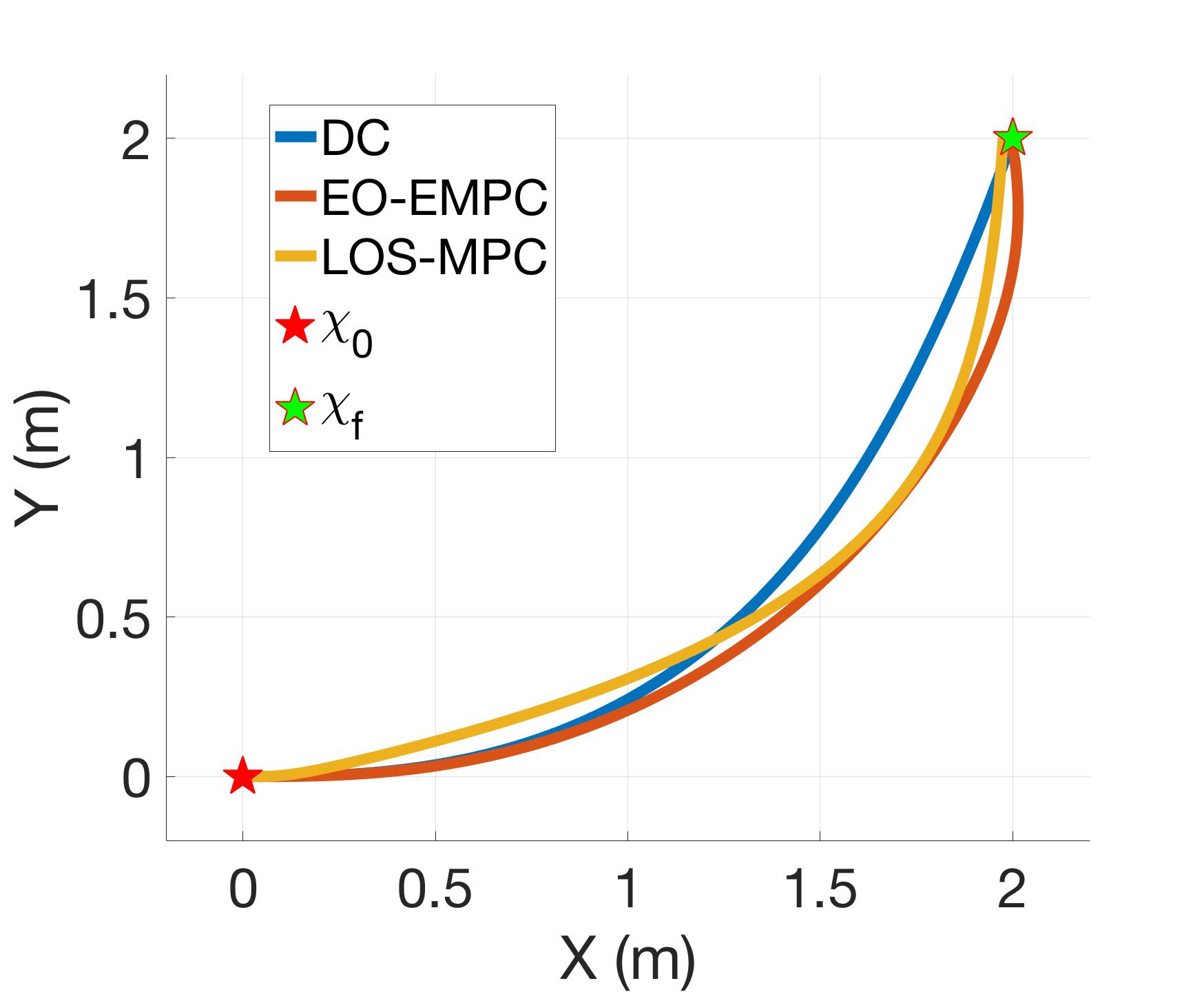} \vspace{-0.3cm}
\caption{Horizontal trajectory comparison ($x_0=0$, $y_0=0$)}  \vspace{-0.2cm}
\label{fig:Ori_Trajectory_Comp} 
\end{figure}

In the first case study, we simulate the vehicle using the three control methods for the scenario considered in Section.~\ref{Section.3B}. The horizontal trajectories resulted from DC, EO-EMPC, and LOS-MPC are shown in Fig.~\ref{fig:Ori_Trajectory_Comp}. As demonstrated in Fig.~\ref{fig:Ori_Trajectory_Comp}, all control algorithms are able to drive the vehicle to the destination. The energy consumption and the travel time spent by all the methods are compared in Table~\ref{table:Performance Comparison (Ori)}. It can be seen from Table~\ref{table:Performance Comparison (Ori)} that the performance of EO-EMPC is very close to that of DC. However, in our simulations, DC takes $1181.04\ s$ of the computation time to obtain the whole trajectory, while EO-EMPC only takes $16.48\ s$. Compared to LOS-MPC, EO-EMPC spends a slightly longer travel time ($+14.59\%$) but consumes much less energy ($-52.89\%$). \vspace{-0.4cm}

\newcommand{\tabincell}[2]{\begin{tabular}{@{}#1@{}}#2\end{tabular}} 
\begin{table}\footnotesize
\begin{spacing}{1.0}
\centering %\vspace{-0.2cm}
\caption{Performance Comparison ($x_0=0$, $y_0=0$)} \vspace{-0.3cm}
\label{table:Performance Comparison (Ori)}
\begin{tabular}{@{}cccc@{}}
\toprule
Method & \tabincell{c}{Travel \\Time (s)} &  \tabincell{c}{Energy \\Consumption (J)}  & \tabincell{c}{CPU Time for\\the whole Trip (s)}  \\ \midrule
DC       &               21.33        &   19.93        &   1181.04    \\ 
EO-EMPC      &          23.40              &    21.91       &     16.48     \\
LOS-MPC      &              20.42          &    46.51      &        6.34            \\ \bottomrule
\end{tabular} \vspace{-0.4cm}
\end{spacing} 
\end{table} 
%

% %
% \begin{figure*} [t!] 
% \centering     %%% not \center
% \subfigure [$x_0=0$, $y_0=-0.5$] {\label{fig:a}
% \includegraphics[width=60mm]{Ori_Traj_Comp_neg05.jpg}}
% \subfigure [$x_0=0$, $y_0=0.5$] {\label{fig:b}
% \includegraphics[width=56mm]{Ori_Traj_Comp_05.jpg}}
% \caption{Horizontal trajectory comparison}
% \label{fig:Per_Trajectory_Comp} 
% \end{figure*} 
% %

%
\begin{figure} [t!]  
\centering    
\subfigure [$x_0=0$, $y_0=-0.5$] {\label{fig:a}
\includegraphics[width=2.5in]{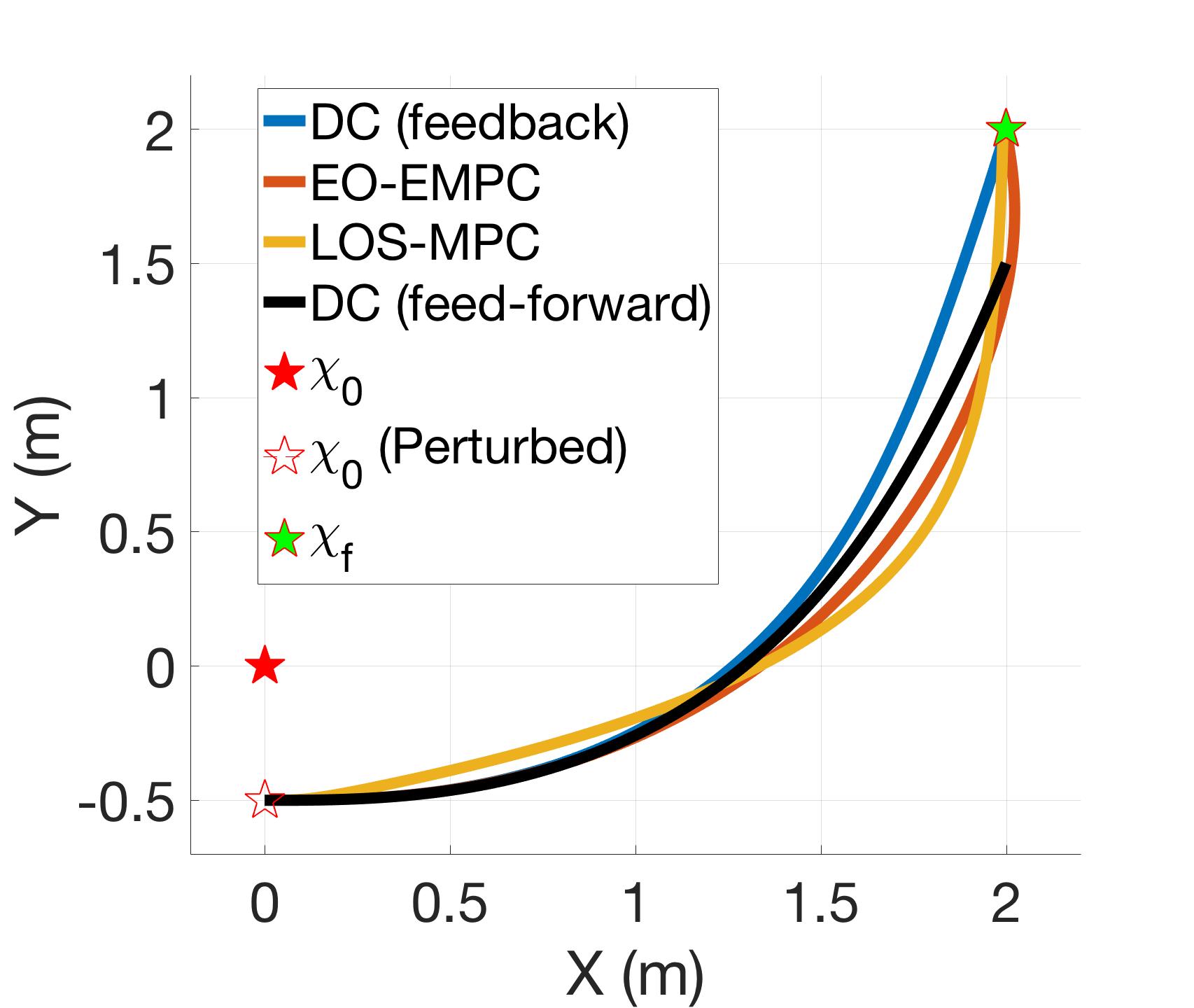}}
\subfigure [$x_0=0$, $y_0=0.5$] {\label{fig:b}
\includegraphics[width=2.5in]{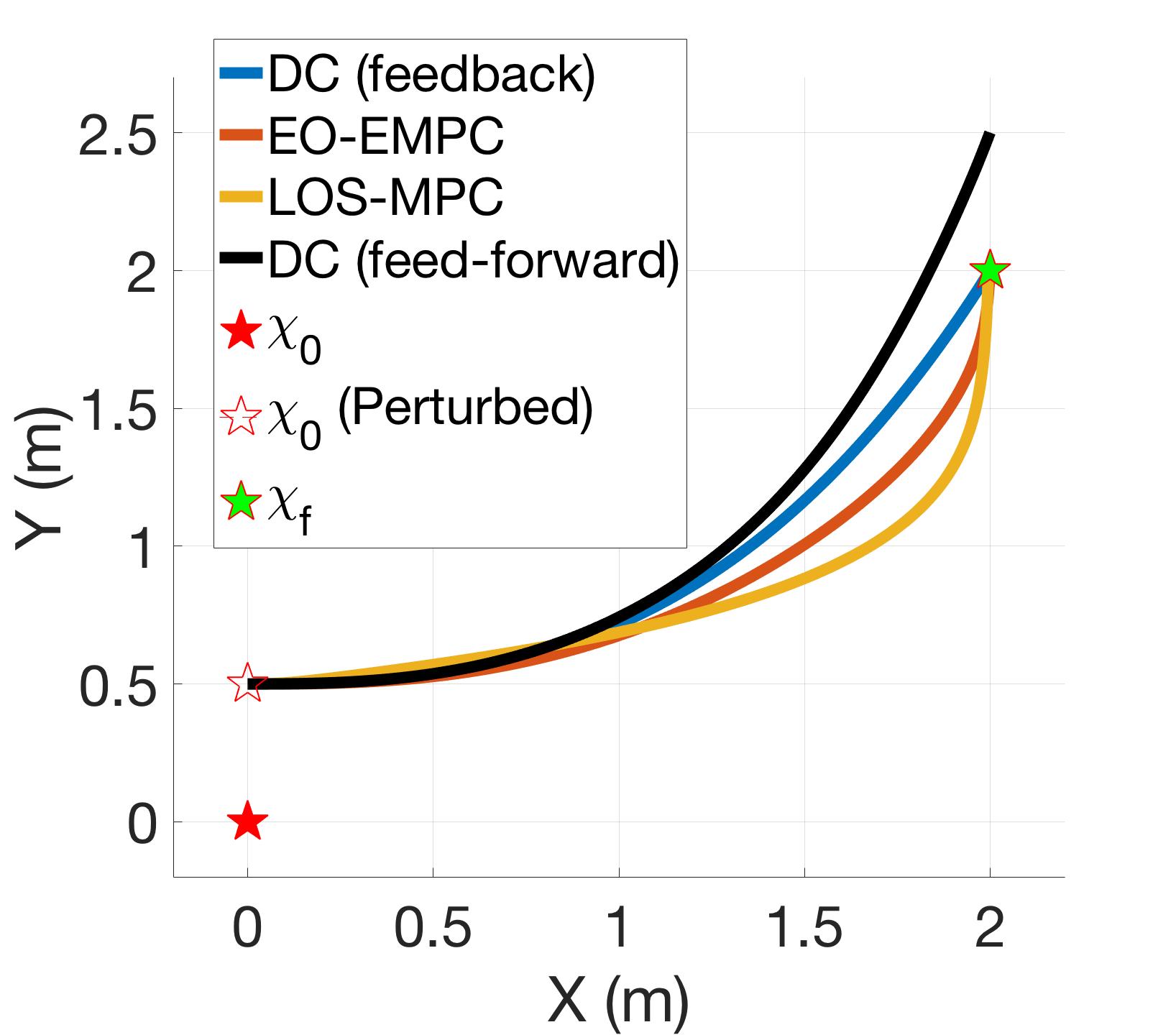}} \vspace{-0.2cm}
\caption{Horizontal trajectory comparison} 
\label{fig:Per_Trajectory_Comp} 
\end{figure}

In order to verify the robustness of EO-EMPC, two additional case studies are carried out with perturbations added to the initial $y$ position.  The $y_0$ for the two cases are perturbed to $-0.5$ and $0.5$, respectively. Two different implementations of DC are employed: DC (feedback) that runs the control input obtained for the perturbed initial position and DC (feed-forward) that runs the control input obtained for the unperturbed initial position. The resulted horizontal trajectories are displayed in Fig.~\ref{fig:Per_Trajectory_Comp}. From Fig.~\ref{fig:Per_Trajectory_Comp}, it can be observed that DC implemented as an open-loop control law suffers from robustness issue against the uncertainty in the initial position. The other three methods reach the final position successfully. A further comparison of the energy consumption is given in Table~\ref{table:Performance Comparison Per}. It can be seen that,  in terms of the energy efficiency, EO-EMPC still achieves a near-optimal solution compared to the DC (feedback) and outperforms the LOS-MPC.

\begin{table}\footnotesize 
\begin{spacing}{1.0}
\centering %\vspace{-0.2cm}
\caption{Performance Comparison (Perturbed $y_0$)} \vspace{-0.3cm}
\label{table:Performance Comparison Per}
\begin{tabular}{@{}cccc@{}}
\toprule
Scenario & Method &  \tabincell{c}{Energy \\Consumption (J)} & \tabincell{c}{Energy \\Reduction (\%)}  \\ \midrule
\multirow{3}{*}{\tabincell{c}{$x_0=0$\\$y_0=-0.5$}  }    &          DC (feedback)       &    22.40       &   -54.13  \\
     &          EO-EMPC           &    23.68      &    -51.51      \\
    &              LOS-MPC       &    48.83       &        --             \\\midrule
\multirow{3}{*}{\tabincell{c}{$x_0=0$\\$y_0=0.5$}}        &          DC (feedback)       &    17.48       &    -58.95 \\
      &          EO-EMPC           &    19.38     &    -54.49     \\
    &              LOS-MPC       &    42.58       &        --             \\ \bottomrule
    
\end{tabular} \vspace{-0.4cm}
\end{spacing}
\end{table} 
%

%%%%%%%%%%%%%%%%%%%%%%%%%%%%%%%%%%%%%%%%%%%%%%%%%%%%%%%%%%%%%%%%%%%%%%%%%%%%%%%%%%%%
%%%%%%%%%%%%%%%%%%%%%%%%%%%%%%%%%%%%%%%%%%%%%%%%%%%%%%%%%%%%%%%%%%%%%%%%%%%%%%%%%%%%
\section{Conclusions} \label{section.6}

In this paper, an EMPC was developed to address the energy-optimal motion control of an AUV in the horizontal plane. The terminal cost of the proposed EMPC was formulated in terms of the energy-to-go to account for the energy consumption beyond the prediction horizon. Two operation modes identified from the DC solution were exploited to partition the energy-to-go into dynamic and static costs, respectively. The combination of the dynamic and static costs was further optimized by modifying the duration of the two operation modes. Simulation results verified the effectiveness of the proposed EO-EMPC with three case studies. The EO-EMPC is able to reduce the energy consumption by $50\%$ compared with the baseline MPC that tracks the reference yaw angle generated with a line-of-sight guidance law. Moreover, compared to the DC method, up to $98\%$ decrease in the computation time is achieved by the EO-EMPC with only $10\%$ difference in the energy efficiency. Future work will be focusing on (i) robustness test under different horizontal motion scenarios, (ii) validation on different AUV platforms, and (iii) extension to 3D motion control problems.

%without significantly increasing the computational complexity
%The combination of the dynamic and static costs was further optimized by modifying the duration of the two operation modes. 
%(i) robustness test under different horizontal motion scenarios, 

%\addtolength{\textheight}{-15cm}   % This command serves to balance the column lengths

%%%%%%%%%%%%%%%%%%%%%%%%%%%%%%%%%%%%%%%%%%%%%%%%%%%%%%%%%%%%%%%%%%%%%%%%%%%%%%%%
\section*{ACKNOWLEDGMENT}
%Dr. Corina Barbalata and Mr. Eduardo Iscar R\"{u}land from DROP Lab at the University of Michigan are gratefully acknowledged for providing the details of the AUV model and their technical comments during the course this study.

The authors thank Dr. Corina Barbalata, and Mr. Eduardo Iscar R\"{u}land from DROP Lab at the University of Michigan for providing details of the AUV model and their technical comments during the course of this study. 

%%%%%%%%%%%%%%%%%%%%%%%%%%%%%%%%%%%%%%%%%%%%%%%%%%%%%%%%%%%%%%%%%%%%%%%%%%%%%%%%

\bibliographystyle{IEEEtran}
\bibliography{Reference}

%%%%%%%%%%%%%%%%%%%%%%%%%%%%%%%%%%%%%%%%%%%%%%%%%%%%%%%%%%%%%%%%%%%%%%%%%%%%%%%%

% \begin{appendices}
% \label{FirstAppendix}

% %\begin{table}
% \begin{scriptsize}
% \centering
% \label{my-label}
% \begin{tabular}{lll}
% $W=200.116 \, N$ & $B = 201.586\, N$ & $m = 20.42\, kg$\\
% $I_{xx} = 0.1205 \,kg\, m^2$  &  $I_{yy} = 0.9431 \,kg\, m^2$  &   $I_{zz} = 1.0061\, kg\, m^2$\\
% $z_g = 0.0018\, m$  &  $l_1 = 0.1694 \, m$            &   $l_2 = 0.2794 \, m$ \\
% $l_3 = -0.0204\,m $  &  $X_{\dot{u}} = -2.042 \, kg$           &     $Y_{\dot{v}} = -32.2013\, kg$   \\
% $Z_{\dot{w}} = -32.2013\, kg$   &  $K_{\dot{p}} = -0.0805\, kg$           &     $M_{\dot{q}} = -2.6834\, kg$  \\
% $N_{\dot{r}} = -2.6834\, kg$   &  $X_{u|u|} = 48.17 \, kg/m$           &     $Y_{v|v|} = 4.11\, kg/m$ \\
% $Z_{w|w|} = 4.11\, kg/m$   &  $K_{p|p|} = 48.17\, kg/m$           &     $M_{q|q|} = 4.11\, kg/m$ \\
% $N_{r|r|} = 4.11\, kg/m$   &  $\rho = 1.025 \,kg/m^3$           &  $r = 0.025 \, m$  
% \end{tabular}
% %\end{table}
% \end{scriptsize}
% \end{appendices}

%

\end{document}